\documentclass[aps,pre,showpacs,amssymb]{revtex4}
\usepackage{amsfonts}
\usepackage{graphicx}
\usepackage{bm}

\def\EE{{\mathbb{E}}}

\begin{document}

\title{Collapse  of a Bose-Einstein condensate induced by fluctuations
of the laser intensity}

\author{J. Garnier}
\affiliation{
Laboratoire de Statistique et Probabilit{\'e}s,
Universit{\'e} Paul Sabatier,
118 Route de Narbonne, 31062 Toulouse Cedex 4, France,\\
Tel. (33) 5 61 55 62 20, 
Fax. (33) 5 61 55 60 89, 
Email: garnier@cict.fr
}

\author{F. Kh. Abdullaev}

\affiliation{Physical-Technical Institute of the Uzbek Academy of Sciences,
         G. Mavlyanov str. 2-b, 700084, Tashkent, Uzbekistan}

\author{B. B. Baizakov}

\affiliation{Physical-Technical Institute of the Uzbek Academy of Sciences,
         G. Mavlyanov str. 2-b, 700084, Tashkent, Uzbekistan\\
and \\
Dipartimento di Fisica "E. R. Caianiello" 
and Istituto Nazionale di Fisica della Materia (INFM), 
Universit{\'a} di Salerno, I-84081 Baronissi (SA), Italy}

\date{\today}

\begin{abstract}
The dynamics of a metastable attractive Bose-Einstein condensate trapped by a 
system of laser beams is analyzed in the presence of small
fluctuations of the laser intensity. It is shown that the condensate
will eventually collapse.
The expected collapse time is inversely proportional
to the integrated covariance of the time autocorrelation function
of the laser intensity and it 
decays logarithmically with the number of atoms. 
Numerical simulations of the stochastic 
3D Gross-Pitaevskii equation confirms analytical predictions for small and moderate 
values of mean field interaction.

\end{abstract}

\maketitle
\section{Introduction}
The experimental realization of Bose-Einstein condensation (BEC)
in dilute atomic gases \cite{davis95,anderson96,bradley97} founded a rapidly
progressing new field of research \cite{dalfovo}. The physical
properties of BECs, which to date comprise eight elements Rb, Na,
Li, H, He, K, Cs, Yb and their isotopes, are predominantly
determined by interatomic forces. Some of the atomic species
($^7$Li, $^{85}$Rb, $^{133}$Cs) possess a negative $s$-wave
scattering length in the ground state and display attractive
interactions. 
The attractive interaction between the atoms causes the collapse
of the BEC so that a stable BEC was not believed to exist \cite{nozieres}.
However, when an external spatial confinement is imposed for instance by
a system of laser beams, a  trapping potential shows up
which can counterbalance the attractive interaction and allows
the formation of a metastable BEC.
When the number of atoms increases, the attractive interaction
becomes stronger and the energy barrier that prevents the 3D BEC 
from collapsing becomes weaker.
To a given trapping potential there corresponds a critical number of atoms
above which the energy barrier vanishes.
The case of a quadratic potential has been studied,
the critical number of atoms has been computed by a variational
approach and by extensive numerical simulations of the Gross-Pitaevskii (GP)
equation, and the results have been checked experimentally
\cite{dalfovo,roberts01,savage03}.

One of the most important aspects of BECs in the regime of
attractive interactions is that they are unstable against
collapse. The collapse shows up as a rapid and strong shrinking
of the condensate at some critical number of atoms, and is
accompanied by significant atomic losses due to many-body
processes \cite{donley}. The collapse is initiated when the
balance of forces governing the size and shape of the condensate
is altered either by internal or external factors.
With respect to spatial and energetic stability the magnetic
traps appear to be better controllable compared to optical traps
\cite{grimm}. On the other hand, due to increasing interest in
far-off resonant laser traps for Bose-condensation of atoms which
are insensitive to magnetic fields \cite{takasu}, the
investigation of different aspects of BEC dynamics in optical
traps is becoming a very relevant subject.
Of particular interest
is the effect of temporal fluctuations
of the laser intensity which in turn involve temporal
fluctuations of the parabolic trapping potential \cite{savard}.
In the present paper we shall consider the BEC dynamics under random
fluctuations of the strength of the parabolic trap potential 
and we shall show that small fluctuations can lead to the eventual 
collapse of the 3D BEC due to a cumulative effect of
stochastic perturbations.
The random fluctuations have all harmonics in their spectrum,
and some of them participate in the parametric resonance
leading to collapse.
This stochastic parametric resonance in the BEC
width oscillations has a rough equivalent particle picture: 
the Kramers' exit problem which is concerned with
noise activated escape from a potential well \cite{hanggi}.

Quantum tunneling (QT) is considered as playing a key role in the
condensate collapse when the number of atoms is close to the critical
number \cite{leggett}.
We shall see that the BEC instability driven by random
fluctuations of the strength of the parabolic trap potential
is all the more
dramatic as the number of atoms is closer to the critical number.
Our consideration thus shows that even weak noise can play a
competitive role in this limit with QT and should be taken into account.
The effect of optical trap noise was previously considered in the
context of stochastic heating of trapped atoms \cite{savard,gehm}.
In a far-off resonant optical trap created by a system of red
detuned lasers the variable trapping potential can be represented
as $V(t,r)=-\alpha |E(t,r)|^2/4$, where $\alpha$ is the atomic
polarizability and $E(t,r)$ is the electric field amplitude. The
dynamics of trapped atoms can be described by the corresponding
Hamiltonian $H=p^2/(2m) + (1/2) m  \omega_0^2 (1+\eta(t))
r^2$, where $\omega_0^2 = k_0^2/m$ is the mean square trap
oscillation frequency, and $k_0$ is proportional to the
time-averaged laser intensity $I_0 \sim |E|^2$. The time
dependent spring constant is determined by fractional fluctuations
of the laser intensity $\eta(t)=(I(t)-I_0)/I_0$ \cite{savard} .
The influence of the fluctuations of the trap potential on
the dynamics of 1D GP type equation has been considered in \cite{ABP}
and the trap and nonlinearity fluctuations in two dimensional BEC in
\cite{ABK,ABG}.

The paper is organized as follows.
In Section \ref{sec:model} we give a description of the model and
apply a variational approach.
In Section \ref{sec:action} we derive the effective dynamics
 of the action-angle variables of the system driven by random perturbations.
Section \ref{sec:smallnl} (resp. \ref{sec:crit}) are devoted to the asymptotic
analysis of the system for small (resp. near-critical) number of atoms.
Finally we check the variational approach and our asymptotic analysis
in Section \ref{sec:num} by performing direct numerical
simulations of the GP equation.

\section{The model and the variational approach}
\label{sec:model}%
We consider the mean-field GP equation for the single-particle
wave function \cite{gross}
\begin{equation}
  \label{eq:schro0}
  i \hbar \psi_t = 
-\frac{\hbar^2}{2m} \Delta \psi + V(t,{\bf r}) \psi + g |\psi|^2 \psi .
\end{equation}
The nonlinear coefficient is
$g= 4 \pi \hbar^2
a_s/m$
where $a_s$ and $m$ are respectively the atomic scattering length and
mass.
The number of atoms is $N=\int |\psi|^2 dx$.
$V$ is the external trapping potential imposed by a system
of laser beams. We consider a harmonic model, but we take into
account temporal fluctuations of the laser intensity which in turn 
induces temporal fluctuations of the quadratic potential
\begin{equation}
  \label{eq:potential}
  V(t,{\bf r}) = \frac{m \omega_0^2}{2} |{\bf r}|^2 [ 1 +  \eta (t) ] .
\end{equation}
For the optical trap $\omega^{2}= \alpha I/(2 m l_{0}^{2})$, 
where $l_{0}$
is the size of the laser beam, $I$ is the intensity, $\alpha$ is a
constant proportional to the 
laser frequency detuning. The random function $\eta(t)$ describes the 
laser intensity fluctuations
$\eta(t) =(I(t)-I_{0})/I_{0}$.
The stationary random process $\eta$ has zero-mean and standard
deviation $\sigma_\eta$.
We shall see in the following that the
standard deviation is not sufficient to predict the collapse of the
BEC, but the coherence time and more generally the power spectral
density of $\eta$ will play a role.

We now cast Eq.~(\ref{eq:schro0}) in a dimensionless form by introducing
the variables
$t'=\omega_0 t$, ${\bf r}' = {\bf r} / r_0$,
$r_0^{-1} = \sqrt{m \omega_0/ \hbar}$,
and $u = \sqrt{4\pi |a_s| r_0^2} \psi$.
This yields the following partial differential equation (PDE)
\begin{equation}
  \label{eq:schro1}
   i u_{t'} = - \frac{1}{2}
\Delta' u + \frac{1}{2} |{\bf r}'|^2 [1+ \eta'(t')] 
u + \sigma_s |u|^2 u ,
\end{equation}
where $\sigma_s = {\rm sgn}(a_s) = \pm 1$ and $\eta'(t') = \eta(t' /
\omega_0)$.
From now on we drop the primes. 
The next step consists in applying the variational approach.
This approximation was first introduced by Anderson \cite{anderson83}
and developed in nonlinear optics \cite{malomed}.
A similar technique was elaborated for the BEC dynamics 
based on the GP equation \cite{perez97}.
The variational ansatz for the wave function of the BEC is chosen as
the Gaussian \cite{dalfovo}
\begin{equation}
  \label{eq:ansatz0}
  u(t,{\bf r}) = A(t) \exp \left( - \frac{|{\bf r}|^2}{2 a(t)^2} +
    \frac{i b(t) |{\bf r}|^2}{2} + i \theta(t) \right) .
\end{equation}
$a(0) r_0$ is the initial BEC rms width in physical variables
$$
a(0) = \frac{\sqrt{2} }{ \sqrt{3} \sqrt{N} r_0} \left( \int  |{\bf r}|^2| 
\psi(t=0,{\bf r})|^2 d^3 {\bf r}
\right)^{1/2}.
$$
The number of atoms is 
$$
N= \frac{\sqrt{\pi} r_0}{4 |a_s|}
 A(0)^2 a(0)^3 = \frac{\sqrt{\pi} r_0}{4 |a_s|} A(t)^2 a(t)^3.
$$
Following the standard procedure \cite{malomed}, 
we substitute the ansatz into the
Lagrangian density generating Eq.(\ref{eq:schro1}) and calculate the
effective Lagrangian density in terms of $A$, $a$, $b$, $\theta$
and their time-derivatives.
The evolution equations for the parameters of the ansatz are then
derived from the effective Lagrangian by using the corresponding
Euler-Lagrange equations.
In particular this approach yields a closed-form 
ordinary differential equation (ODE) for 
the BEC width $a$
\begin{equation}
\label{eq:ode1}
a_{tt} + a (1 + \eta(t) ) =
\frac{1}{a^3} +\frac{\sigma_s P}{a^4} ,
\end{equation}
where $P = \sqrt{2/\pi} N |a_s| /r_0$.
We study in this paper the attractive case ($a_s<0$, $\sigma_s=-1$).
The evolution equation finally reads
\begin{equation}
\label{ode1}
a_{tt} + a (1 +  \eta(t) ) =
\frac{1}{a^3} - \frac{P}{a^4} .
\end{equation}

\section{Action-angle variables}
\label{sec:action}

\subsection{Unperturbed dynamics}
\begin{figure}
\begin{center}
\begin{tabular}{c}
\includegraphics[width=7.cm]{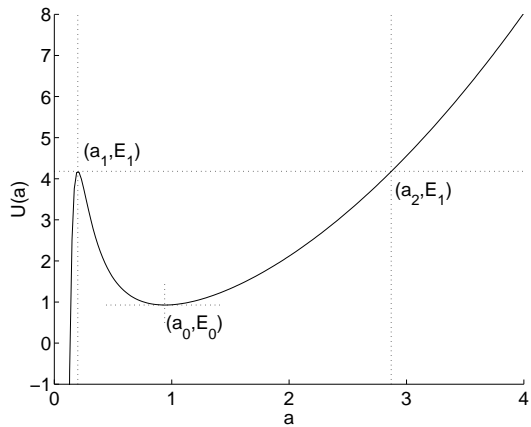}
\end{tabular}
\end{center}
\caption{Potential $U(a)$ for $P=0.2$.
The important points $(a_1<a_0<a_2)$ are also represented.
\label{fig1}
}
\end{figure}

The energy $E$ of the unperturbed BEC is given by:
\begin{equation}
E(t) = \frac{1}{2} a_t^2(t) + {U} (a(t)),
\hspace{0.1in}
{U}(a) = \frac{1}{2} \left( a^2  + \frac{1}{a^2} 
\right) - \frac{P}{3 a^3} .
\end{equation}
In absence of random fluctuations $\eta \equiv 0$ 
the energy $E$ is an integral of motion.
The BEC width obeys a simple dynamics with Hamiltonian
structure
\begin{equation}
\label{mvt0}
H(p,q) = \frac{1}{2} p^2 + {U}(q)
\end{equation}
with $q=a$ and $p=a_t$.
A straightforward analysis \cite{ruprecht,dalfovo} 
shows that if $P<P_c = 4/5^{5/4}
\simeq 0.535$, then the potential $U$ possesses a local minimum
that we shall denote by $a_0$ (see Fig.~\ref{fig1}).
The corresponding ground state has energy $E_0=U(a_0)$.
Below $a_0$ there is the local maximum $a_1$ with energy $E_1=U(a_1)$,
and below $a_1$ the potential decays to $-\infty$.
Above $a_0$ the potential increases to $+\infty$.
It crosses the energy level $E_1$ at $a_2$.

If the initial conditions $(a(0),a_t(0))$
correspond to an energy above $E_1$, or below $E_1$
but $a(0)<a_1$, then the condensate width goes to zero in finite
time which means that the BEC collapses.
On the contrary, if the initial conditions $(a(0),a_t(0))$
correspond to an energy between $E_0$ and $E_1$,
and $a(0)>a_1$, then the orbits of the motion are closed, 
corresponding to periodic oscillations.
In order to explicit the 
periodic structure of the variables $a$ and $a_t$,
we introduce the action-angle variables.
The orbits are determined by the  energy imposed by the initial
conditions:
$$
E =  \frac{1}{2} a_t^2(0) + {U} (a(0)) .
$$
For $E \in (E_0,E_1)$,
we introduce $e_1(E)<e_2(E)$ the extremities
of the orbit of $a$ for the energy $E$:
$$
U(e_1(E)) =  U(e_2(E)) = E .
$$
The action $I$ is defined as a function of the energy $E$ by
\begin{equation}
\label{def:action}
{\cal I} (E)= \frac{1}{2\pi} \oint p dq 
=
\frac{1}{\pi} \int_{e_1(E)}^{e_2(E)} \sqrt{2E-2{U} (b)} db .
\end{equation}
The motion described by (\ref{mvt0})
is periodic, with period
\begin{equation}
\label{def:calt}
{\cal T}(E) 
= \oint \frac{dq}{p} 
=
 2
\int_{e_1(E)}^{e_2(E)} \frac{db}{\sqrt{2E - 2{U}(b)}}  
\end{equation}
or else ${\cal T}(E) = 2 \pi \frac{d{\cal I}}{dE}(E)$.
The angle $\phi$ is defined as a function of $I$ and $a$ by
$$
\phi (E,a) = -\int^{a} \frac{\partial p}{\partial I} dq 
= -\frac{2\pi}{{\cal T}(E)} \int^{a} \frac{db}{\sqrt{2E-2{U}(b)}} . 
$$
The transformation $(E,a) \rightarrow (I,\phi)$
can be inverted to give the functions
${\cal E}(I)$ and ${\cal A}(I,\phi)$.
The BEC width oscillates between the minimum value $e_1(E)$
and the maximum value $e_2(E)$.
The energy $E$ as well as the action $I$ are constant and fixed by the
initial conditions, so the
evolution of the BEC width is governed by
\begin{eqnarray*}
  && a(t) = {\cal A}( {\cal I}(E) , \phi(t)) ,\\
  && \phi(t) = \phi(0) - \frac{2 \pi}{{\cal T}(E)} t .
\end{eqnarray*}

\subsection{Perturbed dynamics}
\label{sec:pert1}%
From now on we assume $\eta \not\equiv 0$ and we denote
by $\sigma_\eta$ the standard deviation of $\eta$.
We investigate the stability of the BEC when the unperturbed
motion is oscillatory.
In particular we aim at studying the collapse time $T_c$
defined as the first time $t$ such that $a(t)=0$.
While the energy of the BEC is below $E_1$, 
the orbit is closed. As soon as the energy reaches the energy level
$E_1$,
the BEC collapses in a time of order $1$ (w.r.t. $\sigma_\eta$).
We shall show that the hitting time for the energy level
$E_1$ is of order $\sigma_\eta^{-2}$, so the collapse time $T_c$
is imposed by the hitting time $T_h$ defined as the first time
$t$ such that $ E(t)=E_1$ or equivalently $I(t)=I_1 := {\cal I}(E_1)$.

In  presence of perturbations, 
the motion of $a$ is not purely oscillatory,
because the energy and the action are slowly varying in time.
We adopt the action-angle formalism, because it allows us to 
separate the fast scale of the locally periodic motion and the slow
scale of the evolution of the action.
Thus, after rescaling $\tau = \sigma_\eta^2 t$ 
the action-angle variables satisfy the differential equations
\begin{equation}
\label{form}
\left\{
\begin{array}{l}
\displaystyle
\frac{ d I}{d\tau} = 
\frac{1}{\sigma_\eta} \eta(\frac{\tau}{\sigma_\eta^2})
h_\phi(I ,\phi) ,\\
\displaystyle
\frac{ d \phi}{d \tau} = - \frac{1}{\sigma_\eta^2}\omega (I) 
- \frac{1}{\sigma_\eta} \eta(\frac{\tau}{\sigma_\eta^2}) 
h_I(I,\phi) ,
\end{array}
\right.
\end{equation}
where $h(I,\phi)= \frac{1}{2} {\cal A}^2(I,\phi)$
and $\omega(I) = \frac{2 \pi }{{\cal T} ( {\cal E} (I))}$
are smooth functions and $h$ is
periodic with respect to $\phi$ with period $2\pi$.
The normalization $\tau = \sigma_\eta^2 t$ has been 
chosen so that the random process $\eta$ appears
with the scales of a white noise in 
the differential equations (\ref{form}).
Applying a standard diffusion-approximation theorem 
\cite{psv}, we get that $(I(t) )_{t \geq 0}$
behaves like a diffusion Markov process
with the infinitesimal generator
$$
{\cal L}_{{I}}= \frac{1}{2} A( {I}) \frac{\partial^2}{\partial  {I}^2}
+ B ( {I})  \frac{\partial}{\partial  {I}}
$$
where
\begin{eqnarray*}
  && A( {I} ) = \frac{1}{\pi} \int_0^{2\pi}
\int_0^{\infty}
h_\phi( {I} ,\phi ) h_\phi( {I} ,\phi - \omega({I}) t) 
\EE[\eta(0) \eta(t) ] dt d\phi ,
  \\
  && B( {I} ) =  \frac{1}{\pi} \int_0^{2\pi} \int_0^{\infty}
h_\phi( {I} ,\phi ) h_{\phi I}( {I} ,\phi - \omega( {I}) t) 
\EE[\eta(0) \eta(t) ] dt d\phi .
\end{eqnarray*}
This means in particular that the probability density function of
${I}(t)$ satisfies the Fokker-Planck equation $\partial_t p = {\cal L}_I^* p$,
$p(t=0,{I})  = \delta({I}-I_0)$, where $I_0$
is the initial action at time $0$ and ${\cal L}_I^*$
is the adjoint operator of ${\cal L}_I$, i.e.
${\cal L}_I^* p=  (1/2) \partial_I^2 \left[ A( {I}) p \right]
- \partial_I \left[ B ( {I}) p \right]$.
Moreover, 
standard results of stochastic analysis allow us to compute
recursively the moments of $T_h$ \cite{feller}.
Denoting  $I_1 = {\cal I}(E_1)$,
the first moment ${\cal \mu}^{(1)} (I)= \EE_I [T_h]$
(the mean value of $T_h$ starting from action $I$ at time $0$)
satisfies
\begin{equation}
\label{feller1}
{\cal L}_I \mu^{(1)} = -1, \ \ \ \ 
\mu^{(1)}(I_1)=0.
\end{equation}
The $n$-th moment ${\cal \mu}^{(n)} (I)= \EE_I [T_h^n]$
satisfies
\begin{equation}
\label{feller2}
{\cal L}_I \mu^{(n)} = -n \mu^{(n-1)}, \ \ \ \ 
\mu^{(n)}(I_1)=0.
\end{equation}
In the following sections we shall apply and discuss these general
results in two different frameworks: small and critical nonlinearity.

\section{Small nonlinearity}
\label{sec:smallnl}%
\subsection{Expansions of the action-angle variables for small nonlinearity}
\begin{figure}
\begin{center}
\begin{tabular}{cc}
{\bf (a)}
\includegraphics[width=7.cm]{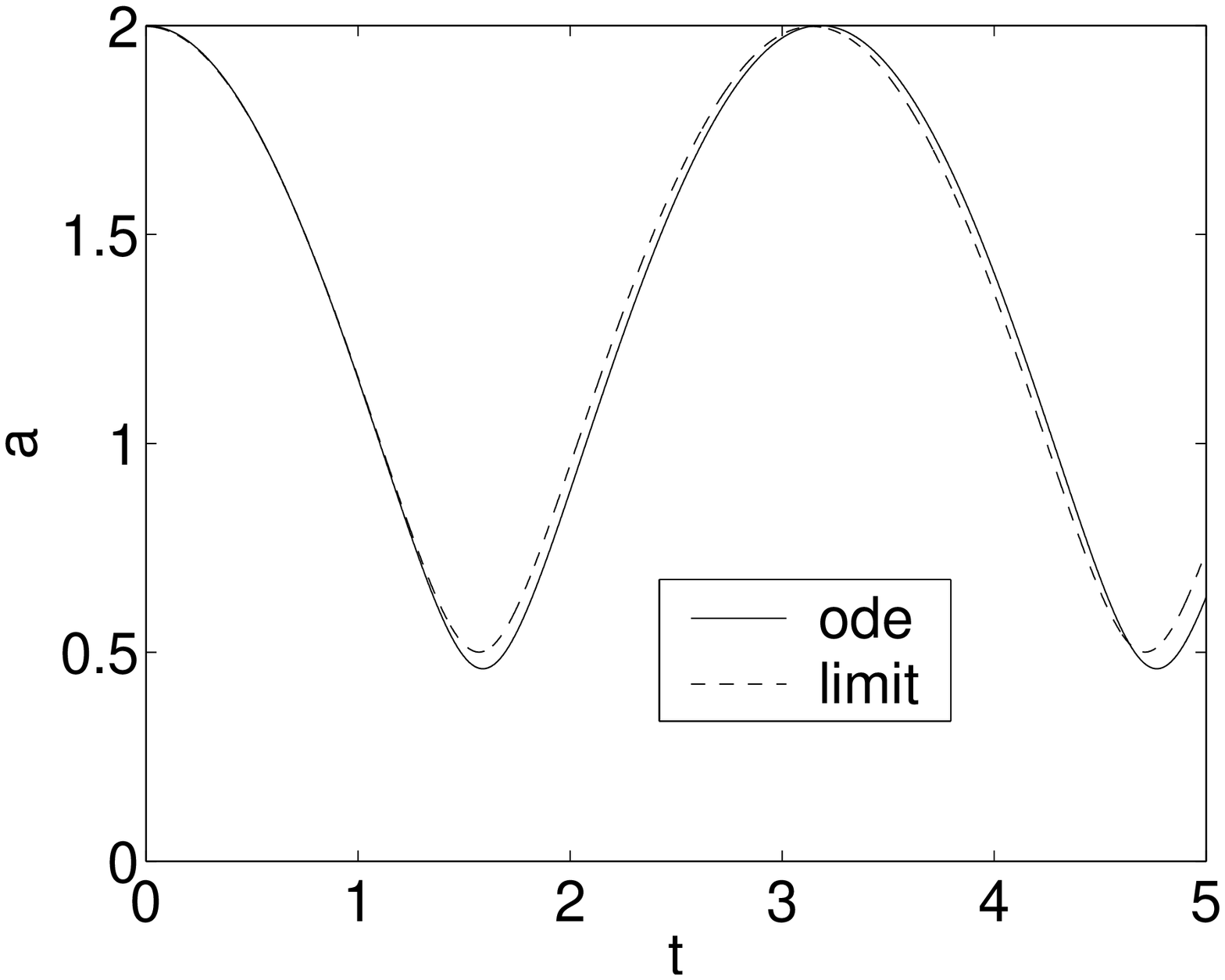}
&
{\bf (b)}
\includegraphics[width=7.cm]{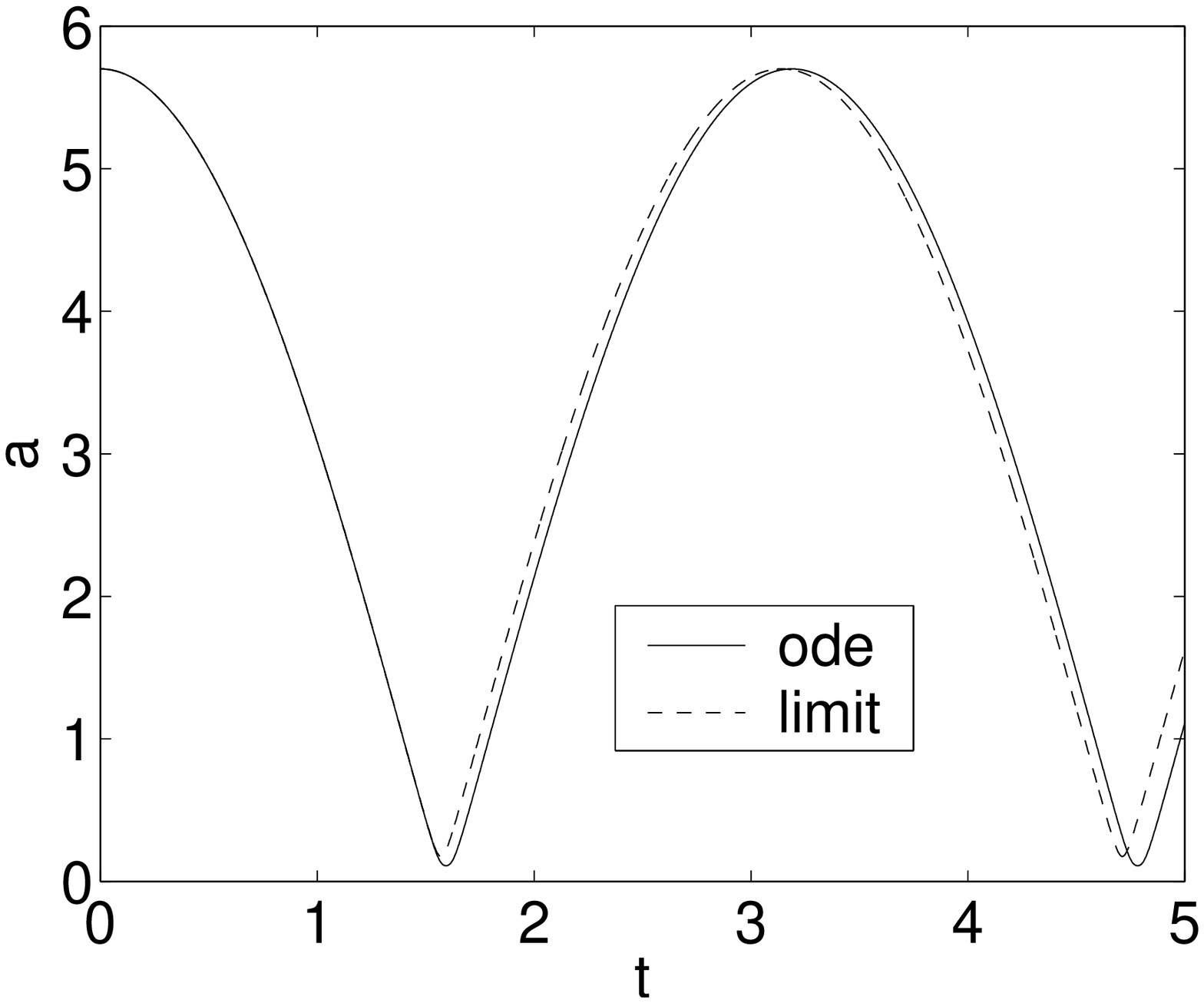}
\end{tabular}
\end{center}
\caption{Unperturbed dynamics of the BEC width.
We assume $P=0.1$, $a_t(0)=0$, $a(0)=2$ (picture a),
$a(0)=5.7$ (picture b).
The second case corresponds to an energy very close to $E_1$.
The results from the resolution of the ODE are compared
with the asymptotic formula (\ref{dyna1}).
\label{fig2}
}
\end{figure}

In this section we assume that $P \ll 1$ which will allow us
to derive simple expressions for the physically relevant quantities.
The points $a_j$ and $E_j$ can be expanded
for small nonlinearity $P$ as
\begin{eqnarray*}
&&
a_0 = 1 
+ O(P), \ \ \ \
a_1 = P +O(P^2) , \ \ \ \
a_2= \frac{1}{\sqrt{3} P} +O(1) ,\\
&& 
E_0 = 1 
+O(P), \ \ \ \
E_1= \frac{1}{6 P^2} + O(\frac{1}{P}) .
\end{eqnarray*}
Note that, as $P$ becomes small, the potential barrier grows like
$P^{-2}$, which shows that the trap looks like a deep 
quadratic external potential.
The functions $h(I,\phi)$ and $\omega(I)$ can also be expanded
for any $\phi $ and $I \leq I_1 = {\cal I}(E_1) = 1/(12 P^2) + 
O( 1/P)$:
\begin{eqnarray*}
  &&h(I,\phi) = \frac{1}{2} + I + \sqrt{I+I^2} \cos(\phi) +O(P) ,\\
  &&\omega(I) = 2 +O(P).
\end{eqnarray*}
Accordingly the unperturbed dynamics of the BEC width for small
$P$ is approximately
\begin{equation}
  \label{dyna1}
  a(t) =   \sqrt{ 1+2I_0+2\sqrt{I_0+I_0^2} \cos(2t)}.
\end{equation}
Figure \ref{fig2} shows that this approximation is indeed 
very good for the orbit $a(t)$
whatever the initial conditions lying in a closed orbit
with energy $< E_1$.

\subsection{Effective equations in presence of perturbations}
In case of small nonlinearity $P \ll 1$, the 
above expansions allow us to derive simple effective equations
for the BEC action in presence of perturbations.
Applying the general results obtained in Section \ref{sec:pert1},
we get that the action $I(t)$ behaves like a diffusion process 
with the infinitesimal generator
$$
{\cal L}_{{I}}= \frac{1}{2} \alpha_c \frac{\partial}{\partial {{I}}}
\left[
({{I}}+{{I}}^2) \frac{\partial}{\partial {{I}}}
\right]
$$
where
$$
\alpha_c = \int_0^\infty \cos(2t) \EE[ \eta(0) \eta(t) ]dt.
$$
The expression of ${\cal L}_{{I}}$ holds true only for 
$ {I}< I_1$.
We can compute the growths of the first moments of 
the action starting from the ground state $I=0$
while $ e^{\alpha_c t} \ll P^{-2}$:
\begin{eqnarray}
\label{momi1}
  && \EE_0[{I}(t)] = \frac{1}{2} e^{\alpha_c t} - \frac{1}{2} ,\\
\label{momi2}
  && \EE_0[{I}(t)^2 ] = \frac{1}{6} e^{3\alpha_c t} -
 \frac{1}{2} e^{\alpha_c t}  + \frac{1}{3} .
\end{eqnarray}
An empirical way to estimate the mean disintegration time
is to look for the time $t_1$ such that $\EE_0[I(t_1)]=I_1$,
where $I_1=1/(12 P^2)$.
From Eq.~(\ref{momi1})  we get $t_1 = (1/\alpha_c) 
\ln[1+1/(6P^2)]$.
This argument is rough because the expectations are ill-placed.
The exact results provided by the rigorous stochastic analysis
confirm that this prediction is not correct.
Integrating Eqs.~(\ref{feller1}-\ref{feller2}) we get 
that the expectation of the disintegration time
starting from the ground state $I=0$ is
\begin{eqnarray}
\label{exptau}
\EE_0[ T_h ] &=&
\frac{2}{\alpha_c} \ln (1 + \frac{1}{12 P^2} )\\ 
\nonumber
&\stackrel{P \ll 1}{\simeq}& \frac{2}{\alpha_c}
\left( - 2 \ln(P)- \ln(12)  \right) ,
\end{eqnarray}
while its variance is
\begin{eqnarray}
\label{vartau}
{\rm Var}_0(T_h) &=& 
\frac{8}{\alpha_c^2} \left[ \ln (1 + \frac{1}{12 P^2} ) 
+ {\rm dilog}(1 + \frac{1}{12 P^2} ) + \frac{1}{2} 
\ln(1  + \frac{1}{12 P^2} )^2 \right]\\
\nonumber
&\stackrel{P \ll 1}{\simeq} 
& \frac{8}{\alpha_c^2}
\left[  -2 \ln(P) - \ln(12) -\frac{\pi^2}{6}    \right],
\end{eqnarray}
where the dilogarithm function is the tabulated function defined as follows:
$$
{\rm dilog}(x) = \int_1^x \frac{\ln(y)}{1-y} dy .
$$
Equations (\ref{exptau}-\ref{vartau}) are the most important results
of this paper. They show that the collapse time varies as 
$\sim \ln (P^{-2})$, while the energy barrier is $\sim P^{-2}$.
In physical variables, the expected collapse time is
$$
\EE_0[ T_c ] = \frac{2}{\omega_0 \alpha} \ln\left(1 +
\frac{\hbar \pi}{24 m \omega_0 a_s^2 N^2} \right), \ \ \ \ 
\alpha= \omega_0 \int_0^\infty \cos(2\omega_0 t) \EE[\eta(0) \eta
(t)] dt.
$$
Taking the experimental data $\omega_{0} = 10$kHz, $N \simeq 5\cdot
10^{3}$, $a_{s}= -5$nm, and $\alpha = 10^{-4}-10^{-5}$,
we obtain the expected collapse time $\approx (1-10)$ seconds.

\subsection{Numerical simulations}
\label{subsec:num1}%
We compare the theoretical predictions 
with numerical simulations of the ODE (\ref{ode1}).
We use a fourth-order Runge-Kutta method for the 
resolution of the ODE.
The random fluctuations
are modeled by a stepwise constant random process:
$$
\eta(t) = \sigma \sum_{j} X_j {\bf 1}_{[ j t_c, (j+1)t_c)}(t) ,
$$
where the $X_j$ are independent and identically distributed 
random variables with uniform distribution  over $(- {1}/2
, {1}/2 )$ and $t_c$ is the coherence time of the laser.
The coefficient $\alpha_c$ is then given by
$$
\alpha_c = \sigma^2 \frac{1-\cos(2 t_c)}{48 t_c}  ;
$$
The first series of simulations were performed with the parameters
$\sigma=0.3$ and $t_c=0.5$.
We investigate different configurations corresponding to different
values of the parameter $P$
starting from $a(0)=1$, $a_t(0)=0$ which is very close to the ground
state.
We have carried $1000$ simulations for each configuration.
The theoretical values for the expected value 
and standard deviation according to formulas 
(\ref{exptau}-\ref{vartau}) are reported in Table \ref{table1}
and compared with the values obtained from averaging of the results
of the numerical simulations.

\begin{table}
\caption{Comparisons between the averages and rms of the 
collapse time obtained from numerical simulations and from theoretical
formulas. Here $\sigma=0.3$ and $t_c=0.5$.
\label{table1}}
\begin{tabular}{|c||c|c|c||c|c|c|}
\hline
 & \multicolumn{3}{c||}{  $\left< \tau \right>$ }
 &\multicolumn{3}{c|}{  ${\rm rms}(\tau)$ } \\
\cline{2-7}
\raisebox{1.5ex}[0cm][0cm]{$P$}
& num & theor & error & num & theor &  error\\
\hline
\hline
$0.05$ &  $4112$ &  $4103$ & $0.2\%$  & $2241$ &  $2335$ &
$4\%$\\
\hline
$0.1$ &   $2585$ &  $2591$ &$0.2\%$  & $1718$ & $1601$ &
$7\%$\\
\hline
$0.2$ &  $1257$ &  $1306$ & $3.5\%$  & $833$ & $865$ &
$4\%$ \\
\hline
$0.3$ &  $586$ &  $760$ & $23\%$  & $407$ & $518$ &
$21\%$ \\
\hline
$0.4$ &  $205$ &  $486$ & $58\%$  & $165$ & $336$ &
$51\%$\\
\hline
\end{tabular}
\end{table}

\begin{figure}
\begin{center}
\begin{tabular}{cc}
{\bf (a)}
\includegraphics[width=7.55cm]{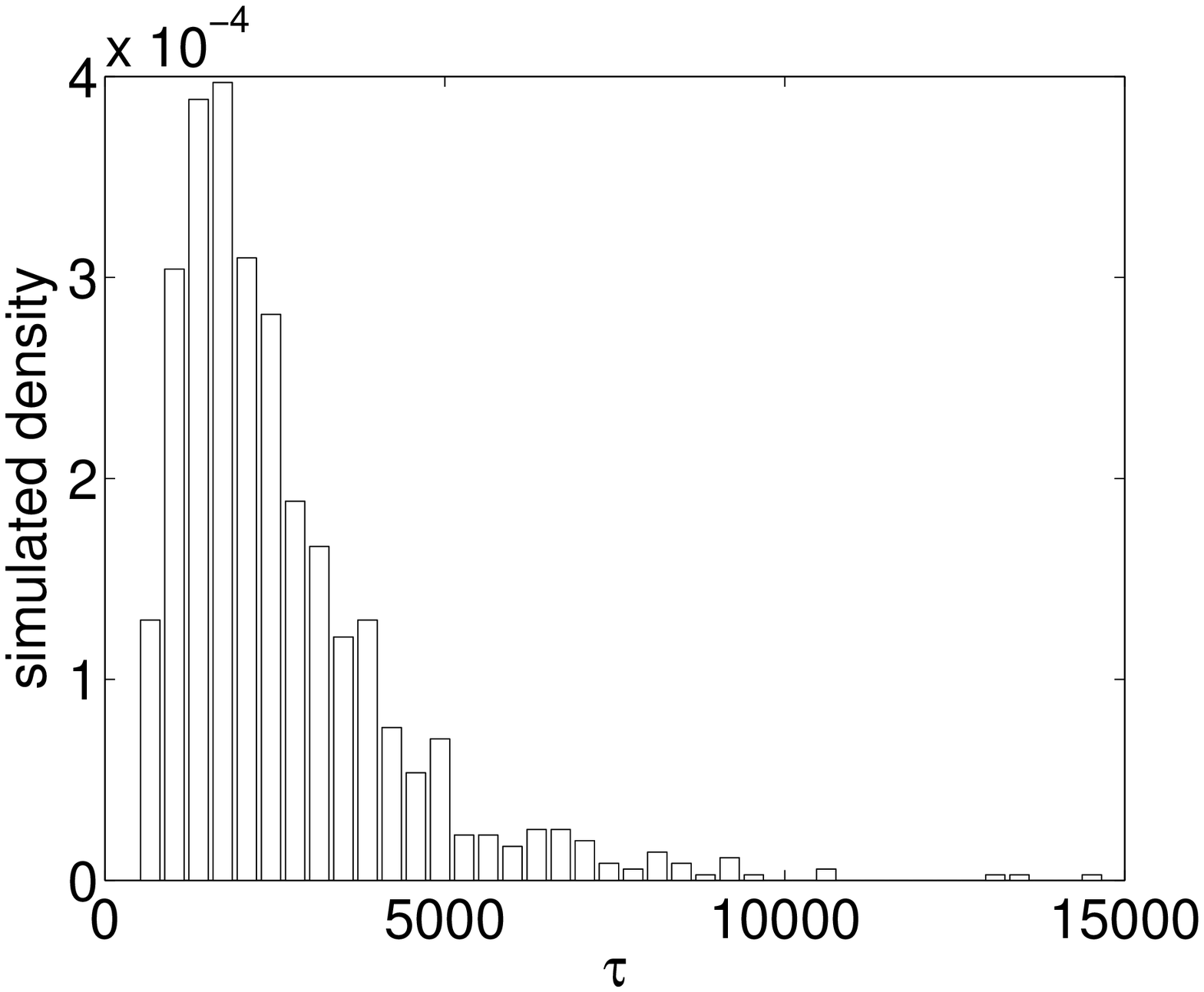}
&
{\bf (b)}
\includegraphics[width=7.cm]{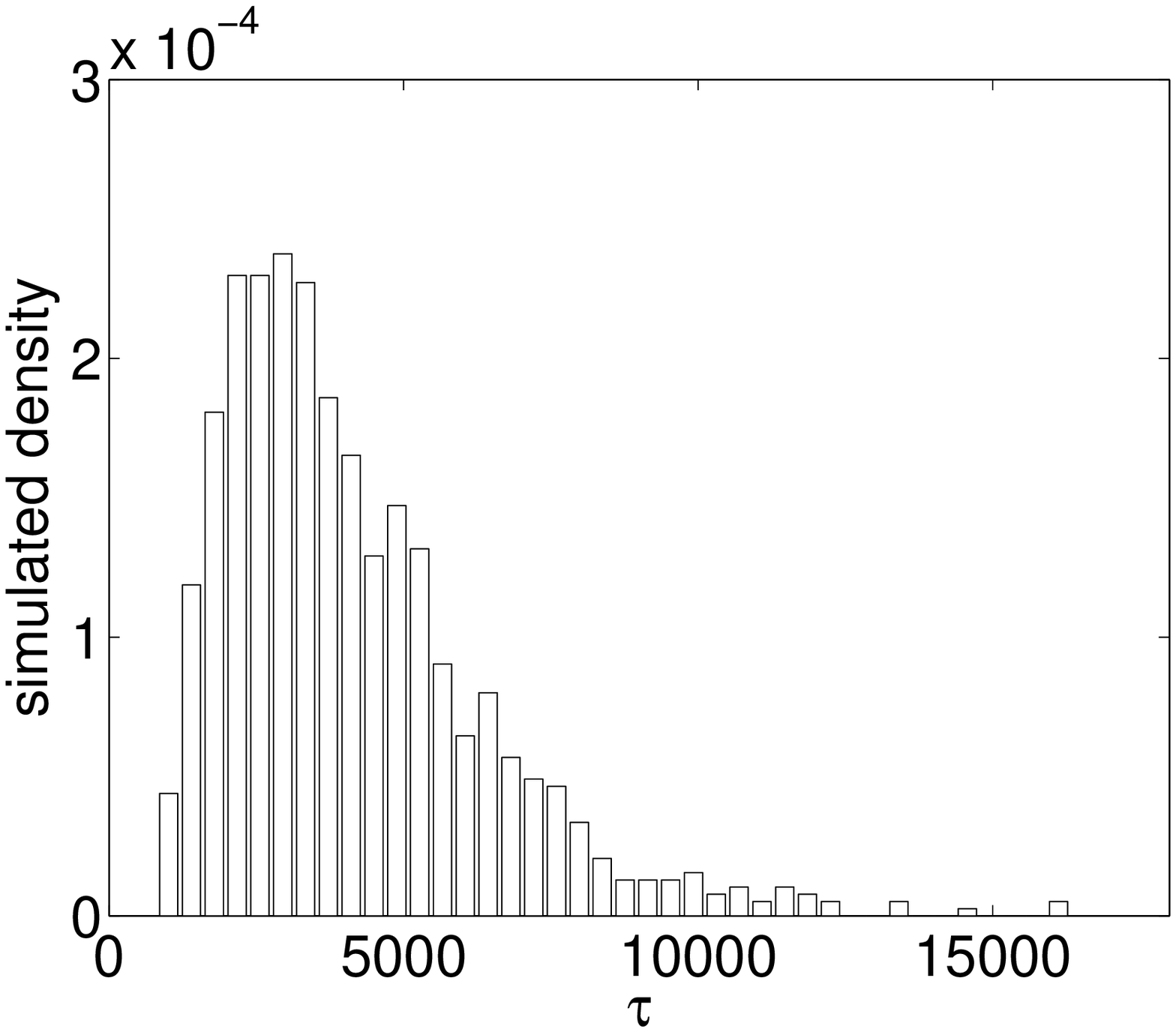}
\end{tabular}
\end{center}
\caption{Histograms of the collapse time 
obtained from series of $1000$ simulations.
Picture a: $P=0.1$. Picture b: $P=0.05$.
\label{fig4}
}
\end{figure}

Note that the statistical formulas are theoretically valid in the
asymptotic framework $P \ll 1$.
The numerical simulations show that they are actually valid
for $P \leq 0.2$.
More exactly, the comparisons between the theoretical predictions
and the numerical simulations
shows excellent agreement for the mean values,
and very good agreement also for the standard deviations.
We also plot in Fig.~\ref{fig4} the histograms of the collapse times
for two series of simulations.

\begin{table}
\caption{Comparisons between the averages and rms of the 
collapse time obtained from numerical simulations and from theoretical
formulas. Here $\sigma=2$ and $t_c=0.5$.
\label{table1b}}
\begin{tabular}{|c||c|c|c||c|c|c|}
\hline
 & \multicolumn{3}{c||}{  $\left< \tau \right>$ }
 &\multicolumn{3}{c|}{  ${\rm rms}(\tau)$ } \\
\cline{2-7}
\raisebox{1.5ex}[0cm][0cm]{$P$}
& num & theor & error & num & theor &  error\\
\hline
\hline
$0.05$ &  $98.6$ &  $92.3$ & $6.4\%$  & $55.5$ &  $52.5$ &
$5.4\%$\\
\hline
$0.1$ &   $63.7$ &  $58$ &$8.5\%$  & $39.1$ & $36.0$ &
$7.9\%$\\
\hline
$0.2$ &  $31.9$ &  $29.4$ & $7.8\%$  & $21.2$ & $19.5$ &
$8.2\%$ \\
\hline
$0.3$ &  $16.1$ &  $17.1$ & $6.5\%$  & $11.4$ & $11.7$ &
$2.6\%$ \\
\hline
$0.4$ &  $6.6$ &  $10.9$ & $65\%$  & $4.9$ & $7.6$ &
$55\%$\\
\hline
\end{tabular}
\end{table}

Finally, in Table \ref{table1b},
we report results with a high level of fluctuations
(namely $\sigma =2$).
The theoretical predictions are still in agreement with
the numerical
simulations for $P \leq 0.3$
with an accuracy of $10 \%$ although the considered
configurations are at the boundary of the validity
of the asymptotic theory.

\section{Critical nonlinearity}
\label{sec:crit}
\subsection{Expansions of the action-angle variables for critical
  nonlinearity}
\begin{figure}
\begin{center}
\begin{tabular}{cc}
{\bf (a)}
\includegraphics[width=7.cm]{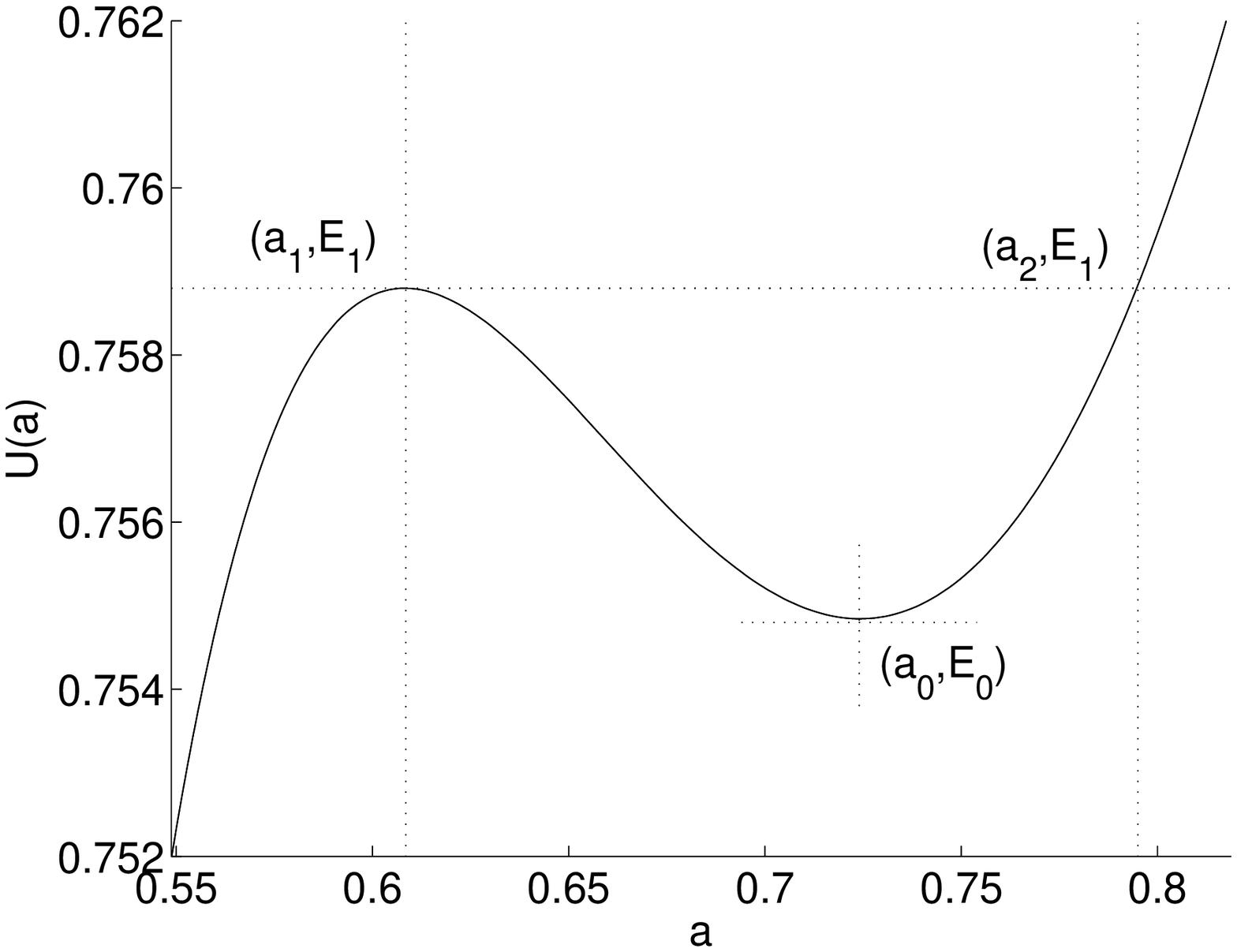}
&
{\bf (b)}
\includegraphics[width=7.cm]{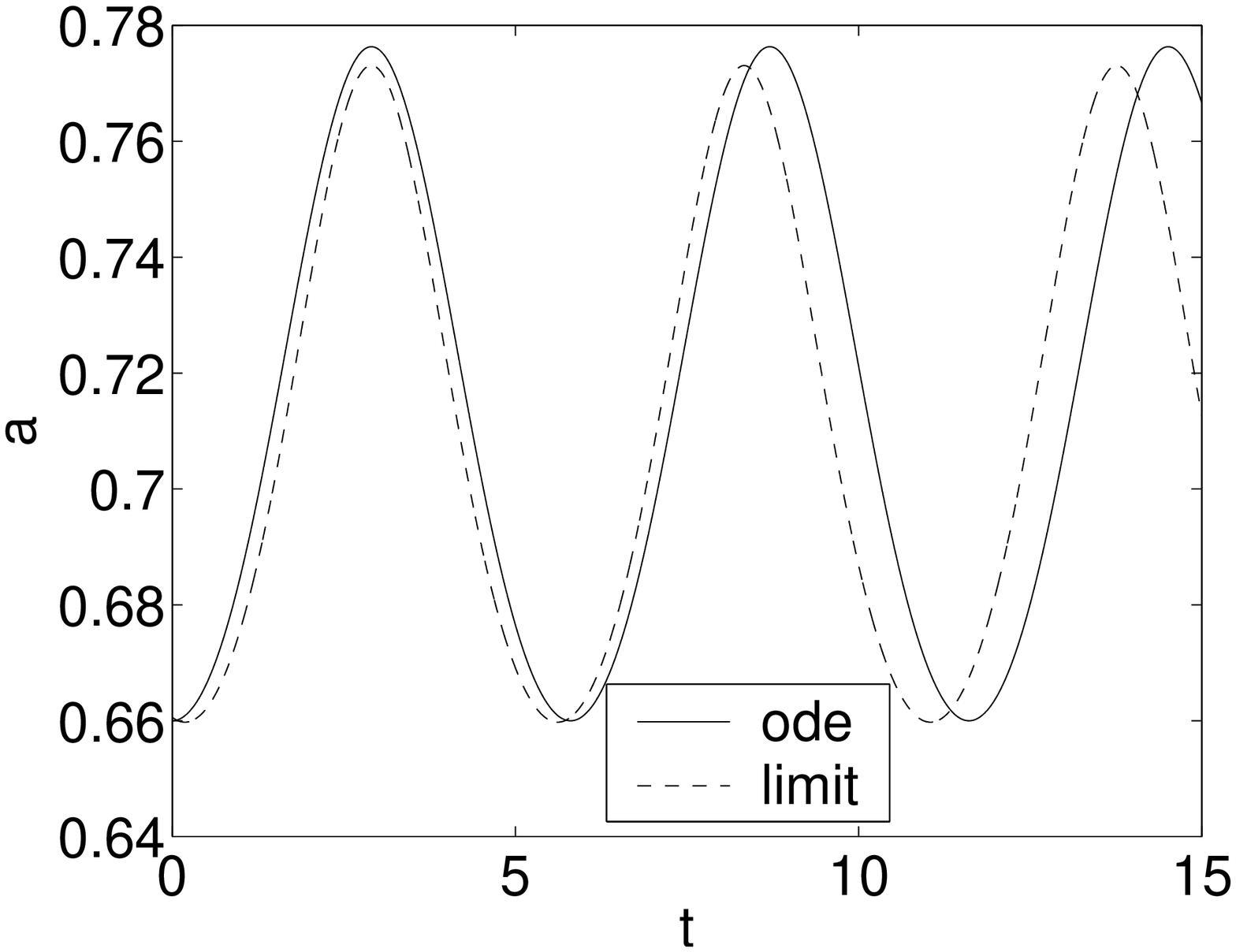}
\end{tabular}
\end{center}
\caption{Picture a: Potential $U(a)$ for $P=P_c-0.01 \simeq 0.525$
($\delta=0.01$).
Picture b:
Unperturbed dynamics of the BEC width.
We assume $a_t(0)=0$, $a(0)=0.66$, $\delta=0.01$.
The results from the resolution of the ODE are compared
with the asymptotic formula (\ref{dyna2}).
\label{fig5}
}
\end{figure}

In this section we address the case where the nonlinear
parameter $P$ is close to the critical value $P_c= 4/5^{5/4}$.
We do so by setting 
$P =P_c - \delta$ and assuming $\delta \ll 1$.
Once again, all quantities can be expanded in powers of
$\delta$. After some algebra, we get
\begin{eqnarray*}
&&  a_j = a_g + 2^{-1/2} 5^{-1/8} \tilde{a}_j \delta^{1/2} + O(\delta)
\mbox{ with } 
a_g = 5^{-1/4}, \ \ \
\tilde{a}_0=1, \ \ \
\tilde{a}_1=-1, \ \ \
\tilde{a}_2=2, \\
&& E_j = E_g + 2^{1/2} 3^{-1} 
5^{7/8} \tilde{E}_j \delta^{3/2}  + O(\delta^2)
\mbox{ with } 
E_g = 3^{-1} 5^{1/2} + 3^{-1} 5^{3/4} \delta ,\ \  \
\tilde{E}_0=-1, \ \ \
\tilde{E}_1=1.
\end{eqnarray*}
More generally,
if $a \in [a_1, a_2]$, then it can be parameterized as
$a = a_g + 2^{-1/2} 5^{-1/8} \delta^{1/2} \tilde{a}$
and the potential at $a$ can be expanded as
$$
U(a) = E_g + 2^{1/2} 3^{-1} 
5^{7/8} \delta^{3/2} \tilde{U}( \tilde{a})   + O(\delta^2) ,
$$
where
$$
\tilde{U}(\tilde{a})  = \frac{1}{2}( \tilde{a}^3 - 3 \tilde{a} ) .
$$
Note that locally (i.e. around $a_g$) the potential presents a 
local minimum at $a_0$ (see Figure \ref{fig5}a),
but the shape of the potential well is very different from the
one observed in the framework $P \ll 1$ (compare with
Fig.~\ref{fig1}).
The width of the well $a_2-a_1$ is of the order $\sqrt{\delta}$ and 
its depth $E_1-E_0$ is of order $\delta^{3/2}$.
The local shape of the potential is given by the cubic function
$\tilde{U}$.

We now consider the action-angle variables.
If $E \in [E_1,E_2)$, then it can be
parameterized as
$E =  E_g + 2^{1/2} 3^{-1} 
5^{7/8} \tilde{E} \delta^{3/2}  $
with
$\tilde{E} \in [\tilde{E}_0, \tilde{E}_1) $.
There exist three solutions $e_3(\tilde{E}) \leq \tilde{a}_1 \leq
e_1(\tilde{E}) \leq e_2(\tilde{E})  \leq \tilde{a}_2$ of the cubic
equation
$\tilde{U}(\tilde{a}) = \tilde{E}$.
$e_1(\tilde{E})$ and $e_2(\tilde{E})$ determine the extremities
of the orbit of the normalized width
$\tilde{a}$ for the normalized energy $\tilde{E}$
in case of unperturbed dynamics.
The cubic equation can be solved:
$$
e_j(\tilde{E}) = 2 \cos \left( \frac{{\rm arccos}(\tilde{E})+2\pi(j-2)}{3}
\right) .
$$
In particular, 
if $E=E_0$ (i.e. $\tilde{E}=\tilde{E}_0$), then 
$e_1(\tilde{E}_0) = e_2(\tilde{E}_0) =1$ (and $e_3(\tilde{E}_0) =-2$),
which corresponds to the ground state $a(t) \equiv a_0$,
or $\tilde{a}(t) \equiv 1$.

The period ${\cal T}(E)$ of the closed orbit 
at energy level $E$, as defined by (\ref{def:calt}),
can be expanded as well.
Introducing 
$$
\tilde{\cal T}(\tilde{E}) = 
\delta^{1/4} 2^{-1/4} 
5^{9/16}{\cal T}( E_g + 2^{1/2}
3^{-1} 5^{7/8} \tilde{E} \delta^{3/2} ) ,
$$ 
we get that 
$\tilde{\cal T}$ is at leading order with respect to $\delta$
a $O(1)$-function that can be expressed in terms of tabulated functions
$$
\tilde{\cal T} (\tilde{E}) = \frac{2 \sqrt{3}}{\sqrt{ e_2(\tilde{E}) -
     e_3(\tilde{E})}} K \left( 
\rho(\tilde{E}  ) 
\right),
$$
where 
$$
\rho(\tilde{E}) = \frac{  e_2(\tilde{E}) -  e_1(\tilde{E})}{ e_2(\tilde{E}) -
  e_3(\tilde{E})}
$$
and $K$ is the complete elliptic integral \cite[p. 590]{abra}.
We then define a normalized action $\tilde{\cal I}(\tilde{E})$ 
for $\tilde{E} \in [\tilde{E}_0,\tilde{E}_1]=[-1,1]$ by
$$
\tilde{\cal I} (\tilde{E} ) = 
\frac{1}{2\pi} \int_{-1}^{\tilde{E}} \tilde{\cal T} (s)ds .
$$
The function 
$\tilde{\cal I} :
[\tilde{E}_0 , \tilde{E}_1] \rightarrow [0 , \tilde{I}_1] $
is invertible.
Its inverse is denoted by
$\tilde{\cal E} :
[0 , \tilde{I}_1]\rightarrow
[\tilde{E}_0 , \tilde{E}_1] $
where $\tilde{I}_1 = 18/(5 \pi)$.
It is plotted in Fig.~\ref{fig:crit1}a.
We can see that $\tilde{\cal E}$ is roughly linear.
Similarly we can define the angle ${\cal \phi}( \tilde{E} ,
\tilde{a})$
and its inverse $\tilde{\cal A} ( \tilde{I} , \phi)$.
The function $\tilde{\cal A} :
[0 , \tilde{I}_1] \times [0,2\pi) \rightarrow
[\tilde{a}_1 , \tilde{a}_2] $
can be expressed in terms of Jacobian elliptic functions
\begin{equation}
\tilde{\cal A} ( \tilde{I} , \phi)=
{e}_1 ( \tilde{\cal E}(\tilde{I}) )
+
\left[ {e}_2 ( \tilde{\cal E}(\tilde{I}) )
-
{e}_1 ( \tilde{\cal E}(\tilde{I}) ) 
\right]
{\rm sn}^2 \left( \frac{K( \rho (\tilde{\cal E}(\tilde{I}))}{\pi} \phi,
\rho (\tilde{\cal E}(\tilde{I})) \right) ,
\end{equation}
where sn is the Jacobian sinus 
\cite[p. 589]{abra}.
In absence of perturbation the action is preserved and
the closed orbit of $\tilde{a}(t)$ for a normalized action 
$\tilde{I} \in [0, \tilde{I}_1)$
is given by
\begin{equation}
\label{dyna2}
\tilde{a}(t)  =
\tilde{\cal A} \left(\tilde{I}, \phi(t) \right) \ \mbox{ with } \
\phi(t) =- \delta^{1/4} 2^{-1/4} 3^{-1/2} 5^{9/16}
\frac{2\pi}{\tilde{\cal T}(\tilde{\cal E}(\tilde{I}))} .
\end{equation}
The true orbit is $a(t) = a_g + 2^{-1/2} 5^{-1/8} \delta^{1/2} 
\tilde{a}(t)$.
Figure \ref{fig5}b shows that this approximation (derived
in the asymptotic framework $\delta \ll 1$) is indeed 
reasonably good.

\subsection{Effective equations in presence of perturbations}
\begin{figure}
\begin{center}
\begin{tabular}{cc}
{\bf (a)}
\includegraphics[width=7.cm]{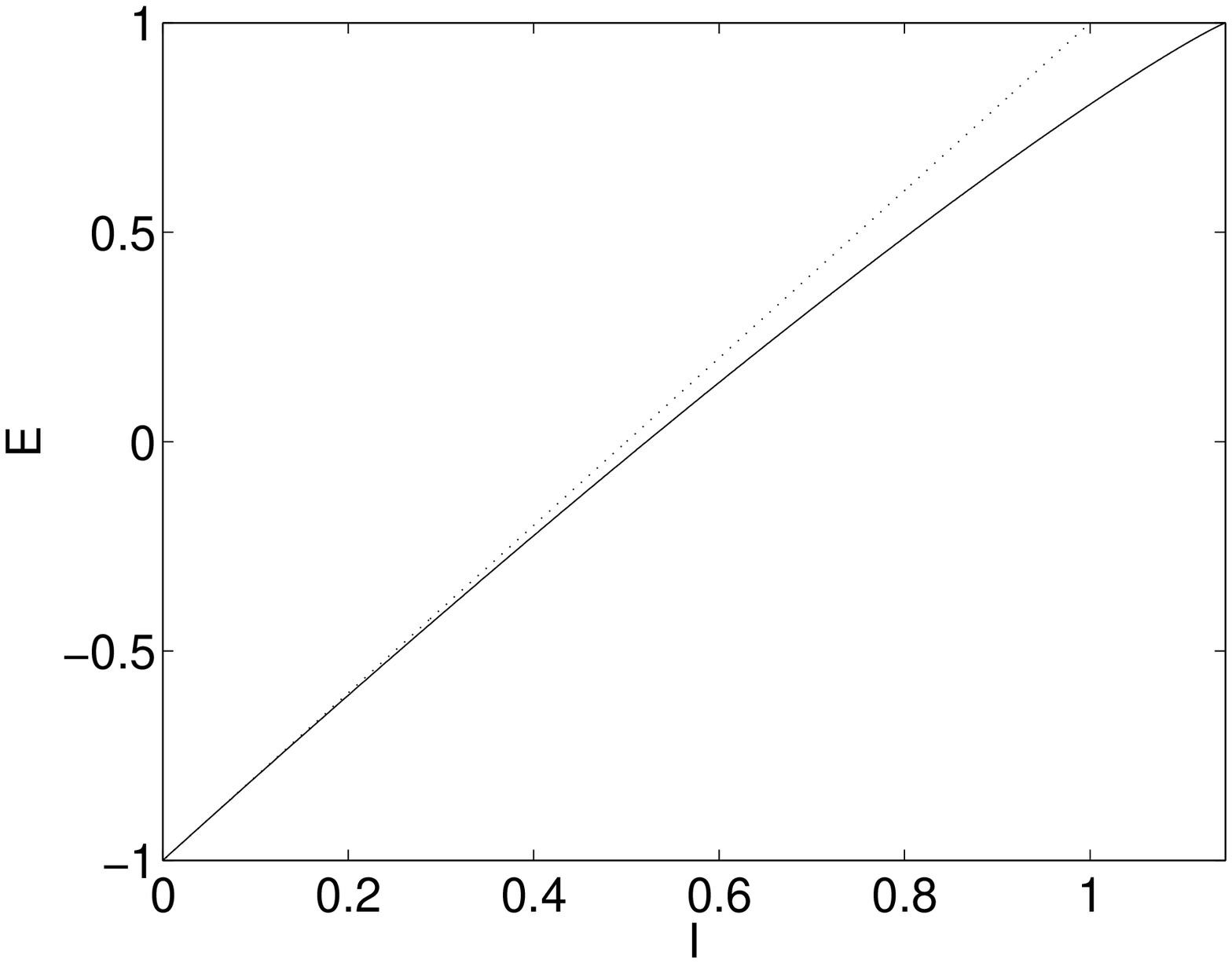}
&
{\bf (b)}
\includegraphics[width=7.cm]{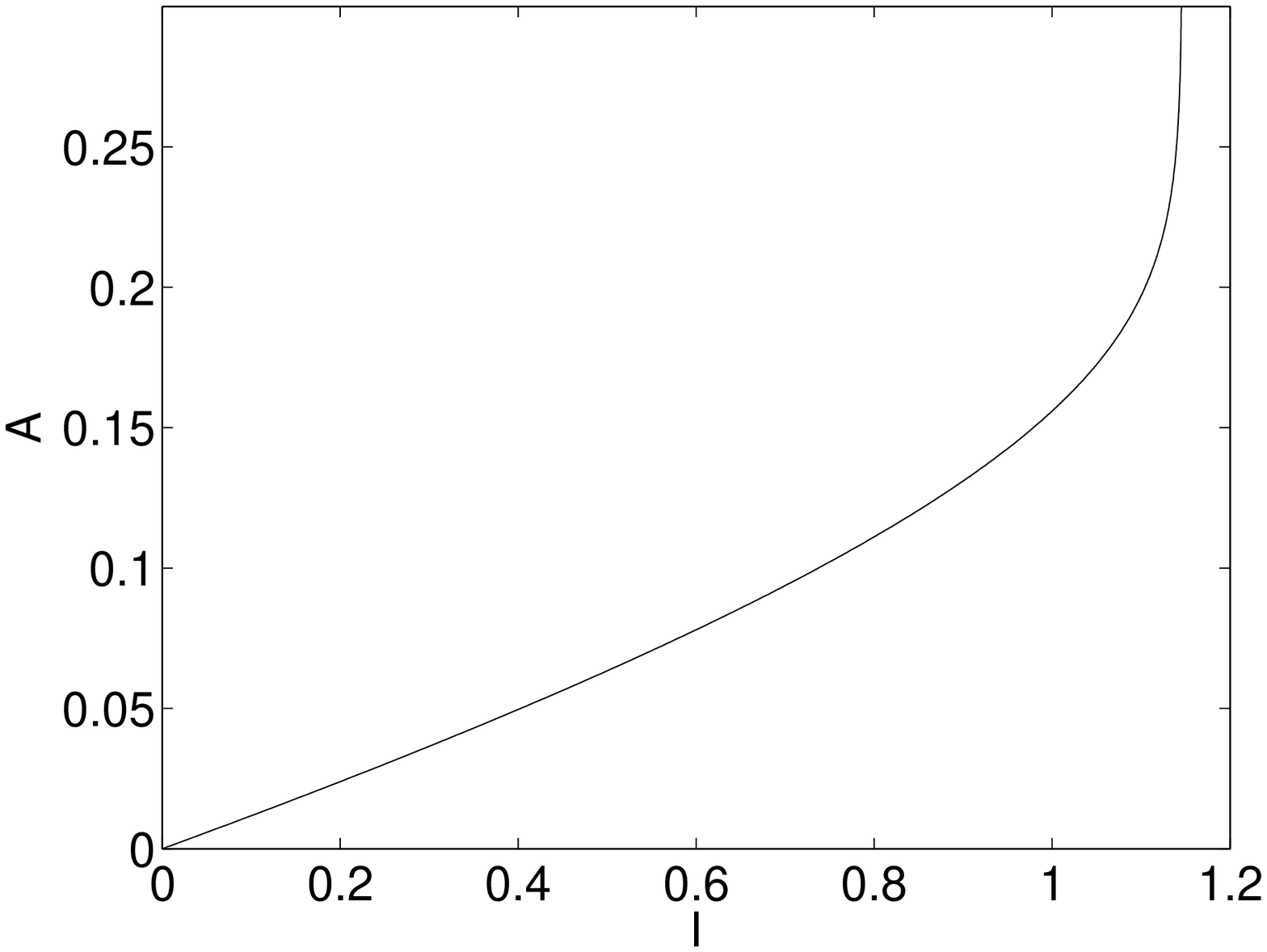}
\end{tabular}
\end{center}
\caption{Functions $\tilde{I} \mapsto \tilde{\cal E}(\tilde{I})$
(picture a) and $\tilde{I} \mapsto \tilde{A}(\tilde{I})$ (picture b).
\label{fig:crit1}
}
\end{figure}

Following the strategy presented in Section \ref{sec:pert1},
we introduce the normalized action-angle variables
so that 
$\tilde{E}(t) = \tilde{\cal E}(\tilde{I}(t)) $ and
$\tilde{a}(t) = \tilde{\cal A}(\tilde{I}(t),\phi(t)) $.
While the energy of the BEC is below $E_1$, 
the orbit is closed. As soon as the energy reaches the energy level
$E_1$,
the BEC collapses in a time of order $1$ (w.r.t. $\sigma_\eta$).
We shall show that the hitting time for the energy level
$E_1$ is of order $\sigma_\eta^{-2}$, so the collapse time $T_c$
is imposed by the hitting time
$T_h$ defined as the first time $t$ such that $\tilde{I}(t)= \tilde{I}_1 $.
Here we rescale $\tau = \sigma_\eta^2 \delta^{-3/2} t$.
This normalization is chosen so that the random process $\eta$ appears
with the scales of a white noise in 
the differential equations
$$
\left\{
\begin{array}{l}
\displaystyle
\frac{ d \tilde{I}}{d\tau} = 
\frac{1}{\varepsilon} \eta(\frac{\tau}{\varepsilon^2})
\tilde{h}_\phi(\tilde{I},\phi) ,\\
\displaystyle
\frac{ d \phi}{d\tau} = - \frac{\delta^{1/4}} 
{\varepsilon^2} 2^{-1/4} 5^{9/16}  
\tilde{\omega} (I) 
- \frac{1}{\varepsilon} \eta(\frac{\tau}{\varepsilon^2}) 
\tilde{h}_{\tilde{I}} (\tilde{I},\phi) ,
\end{array}
\right.
$$
where $\varepsilon=\sigma_\eta \delta^{-3/4}$,
$\tilde{h}(\tilde{I},\phi)=2^{-5/4} 3\ 5^{-11/16} 
\tilde{\cal A}(\tilde{I},\phi)$,
and
$\tilde{\omega}( \tilde{I} ) = \frac{2 \pi }{\tilde{\cal T} 
( \tilde{\cal E} (\tilde{I}))}$.
Note once again that $\tilde{h}$ and $\tilde{\omega}$
are smooth functions, and $\tilde{h}$  is periodic with respect to
$\phi$ with period $2  \pi$.
By applying a diffusion approximation theorem \cite{psv},
we get that $(\tilde{I}(t) )_{t \geq 0}$
behaves like a diffusion Markov process with 
the infinitesimal generator
$$
{\cal L}_{\tilde{I}}= \alpha \delta^{-3/2}
\frac{\partial}{\partial  \tilde{I}}
\tilde{A}( \tilde{I}) \frac{\partial}{\partial  \tilde{I}} 
$$
where
\begin{eqnarray*}
 \tilde{A}( \tilde{I} ) 
&=& 2^{-1/2} 3^2 5^{-11/8} 
\left[ \tilde{e}_2 ( \tilde{\cal E}(\tilde{I}) )
-
\tilde{e}_1 ( \tilde{\cal E}(\tilde{I}) ) 
\right]^2 \frac{K( \rho (\tilde{\cal E}(\tilde{I}))}{\pi^2}
\int_0^{K( \rho (\tilde{\cal E}(\tilde{I}))}
{\rm cn}^2
{\rm dn}^2
{\rm sn}^2 
\left(s,
\rho (\tilde{\cal E}(\tilde{I})) \right) ds ,
  \\
\alpha &=& \int_0^{\infty} \EE[\eta(0) \eta(t) ] dt  ,
\end{eqnarray*}
and dn and cn are two tabulated elliptic functions
\cite[p. 589]{abra}.
The conditions ensuring the diffusion-approximation are
$\delta \ll 1$,  $\sigma_\eta^2 \ll \delta^{3/2}$.
The diffusion coefficient $\tilde{A}(\tilde{I})$ is plotted in
Fig.~\ref{fig:crit1}b. 

Using the results reported in Section \ref{sec:pert1}
we get the following recursive relation ($n \geq 1$)
for the moments of the hitting time $T_h$
\begin{eqnarray}
\label{sysmom}
 && \EE_{\tilde{I}} [T_h^n] = \frac{n \delta^{3/2} }{\alpha}
  \int_{\tilde{I}}^{\tilde{I}_1} 
\frac{\int_0^x \EE_y[T_h^{n-1}] dy}{\tilde{A}(x)}  dx ,
\end{eqnarray}
where $\tilde{I}_1 =18/(5 \pi)$.
In dimensional variables, the result reads as follows. 
Starting from the ground state $a_0$, 
the expected value of the collapse time is
\begin{equation}
\label{exptau2}
\EE_0 [ T_c ] = \frac{
\left(P_c - P \right)^{3/2} }
{\omega_0^2 \alpha} C_1 , 
\end{equation}
where $C_1$ is the constant
$C_1 = \int_{0}^{\tilde{I}_1}
\frac{x}{\tilde{A}(x)} dx$.
By a numerical integration using {\sc Matlab} we have found $C_1 \simeq 8.5$.
More generally, we have
\begin{equation}
\label{vartau2}
\EE_0 [ T_c^n ] = \frac{\left(P_c - P
\right)^{3n/2} }{\omega_0^{2n} \alpha^n} C_n,
\end{equation} 
where $C_n$ are constants obtained recursively from Eq.~(\ref{sysmom}).
By a numerical integration we have found $C_2 \simeq 110$.

\subsection{Numerical simulations}
We compare the theoretical predictions 
with numerical simulations of the ODE (\ref{ode1}).
We use the same model as in Section \ref{subsec:num1}
with the parameters $\sigma=0.025$ and $t_c=0.5$.
We report in Table \ref{table2}
the theoretical values for the expected value 
and standard deviation according to formulas 
(\ref{exptau2}-\ref{vartau2}) as well as 
the values obtained from averaging of the results
of the numerical simulations.
The statistical formulas are theoretically valid in the
asymptotic framework $\delta (=P_c-P) \ll 1$.
The numerical simulations show that they are actually valid
for $\delta \leq 0.03$.

\begin{table}
\caption{Comparisons between the averages and rms of the 
collapse time obtained from numerical simulations and from theoretical
formulas.
\label{table2}}
\begin{tabular}{|c|c||c|c|c||c|c|c|}
\hline
 &
 & \multicolumn{3}{c||}{  $\left< \tau \right>$ }
 &\multicolumn{3}{c|}{  ${\rm rms}( \tau )$ } \\
\cline{3-8}
\raisebox{1.5ex}[0cm][0cm]{$P$}
&
\raisebox{1.5ex}[0cm][0cm]{$\delta$}
& num & theor & error & num & theor &  error\\
\hline
\hline
$0.525$ & $0.01$ & $651$  &  $653$ & $0.3\%$   & $447$ &  $472$ &
$5 \%$\\
\hline
$0.515$ & $0.02$ &  $1754$ &  $1846$ & $5\%$  & $1240$ &  $1334$ &
$7.5\%$\\
\hline
$0.505$ & $0.03$ &  $3175$ &  $3392$ & $7\%$  & $2217$ & $2451$ &
$10.5\%$\\
\hline
$0.495$ & $0.04$ &  $4673$ &  $5222$ & $11.5\%$  & $3107$ & $3775$ &
$21.5\%$ \\
\hline
\end{tabular}
\end{table}

\section{Validation of the variational approach}
\label{sec:num}
The analysis carried out in this paper is based on the variational
approach using a Gaussian ansatz.
The Gaussian ansatz for the study of static and dynamic properties of
trapped gases has been widely used (see for instance
\cite{perez97,shi97,stoof97,parola98,abdullaev01}).
The variational principle is shown in these papers
to be a simple Lagrangian-based method that
gives
reasonable accurate ordinary differential equations approximations
to the true dynamics for the solution of the GP equation.
This method merely assumes Gaussian pulse shapes containing a fixed number 
of free parameters and the Lagrangian form of the partial differential
equation is used to obtain the parameter evolution equations.
However it is a questionable approach
because it is based on the a priori assumption that the solution of
the PDE has a form which remains very close to the chosen ansatz.
Accordingly it has to be checked carefully by full
numerical simulations of the PDE.

\begin{figure}
\begin{center}
\begin{tabular}{cc}
{\bf (a)}
\includegraphics[width=7.cm]{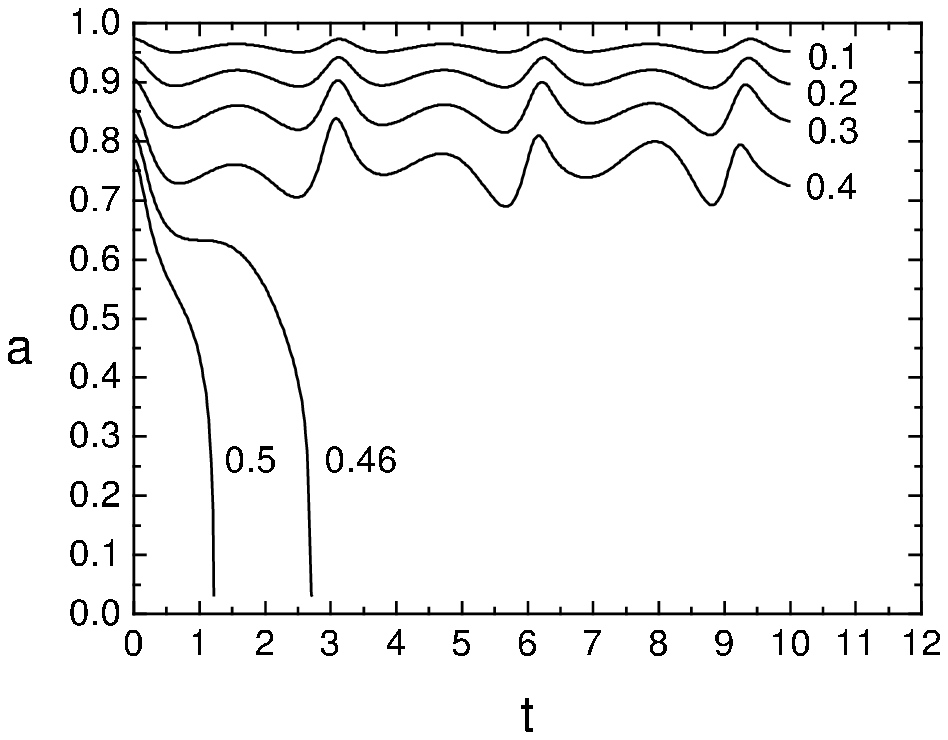}
&
{\bf (b)}
\includegraphics[width=5.7cm]{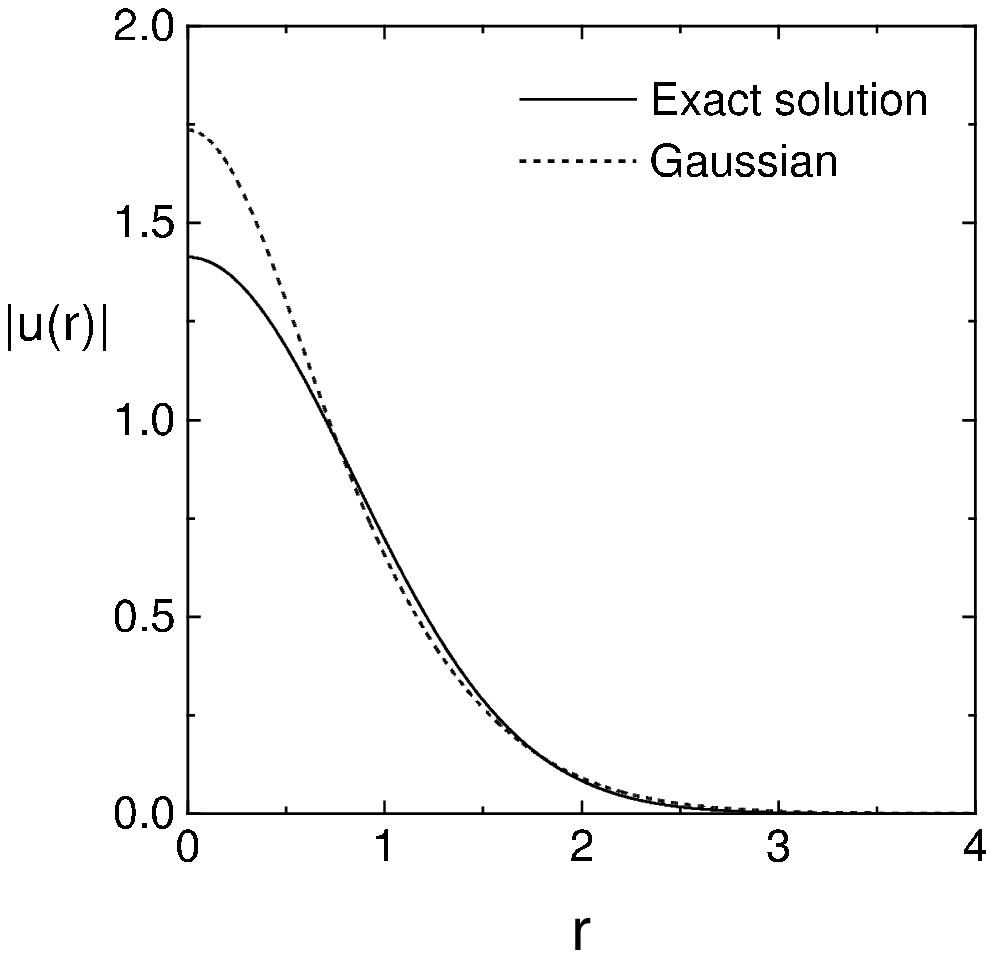}
\end{tabular}
\end{center}
\caption{
\label{fignum1}
Picture a:
Width of the BEC for an initial Gaussian waveform with parameters
corresponding to a stationary point of the potential $U(a)$.
The oscillations are insignificant for small values of $P$, and
become important when $P$ approaches the critical value $P_c =0.459$.
At overcritical $P$ the waveform rapidly shrinks ($a \rightarrow 0$),
i.e. the BEC undergoes collapse.
Picture b: Exact solution of the 3D GP equation (solid line) compared with 
the Gaussian approximation with the same number of atoms and $P=0.44$.}
\end{figure}

Numerical simulations of the stochastic GP equation with spherically
symmetric trap is performed by Crank-Nicholson scheme. The
absorbing boundary condition is employed to imitate the infinite
domain size. This technique allows to prevent re-entering of
linear waves emitted by the condensate under perturbation into the
integration domain.
We have first checked the variational approach
for the unperturbed system.
We have done so by inserting the Gaussian waveform with the 
amplitude and width corresponding to a stationary point
(as predicted by the variational approach)
as an initial condition into the PDE (\ref{eq:schro1}).
We have let the solution evolve in time and we have plotted 
the results in Fig.~\ref{fignum1}a.
As can be seen the Gaussian ansatz is a good approximation 
when $P$ is not close to the critical value $P_c$.
Actually we have found numerically that the critical value for
the existence of the BEC is not $P_c= 0.535$,
as predicted by variational approximation,
but $P_c = 0.459$.
For $P$ very close to the real value of $P_c$,
the Gaussian ansatz  substantially
deviates from the exact solution of the 3D GP
equation, as shown in Fig.~\ref{fignum1}b.

\begin{figure}
\begin{center}
\begin{tabular}{cc}
{\bf (a)}
\includegraphics[width=7.cm]{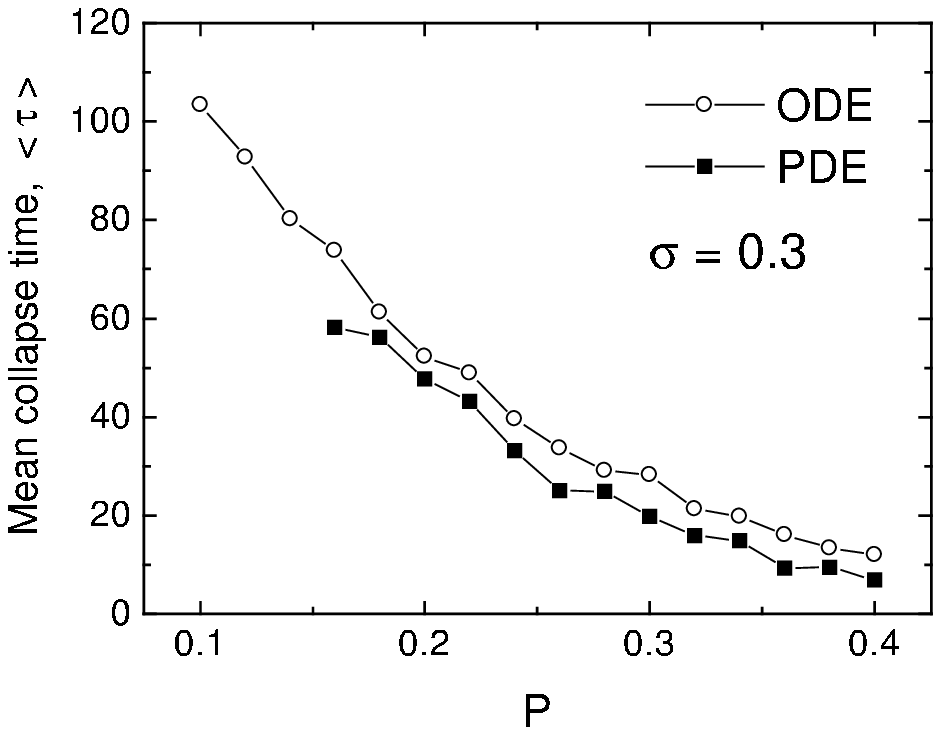}
&
{\bf (b)}
\includegraphics[width=7.cm]{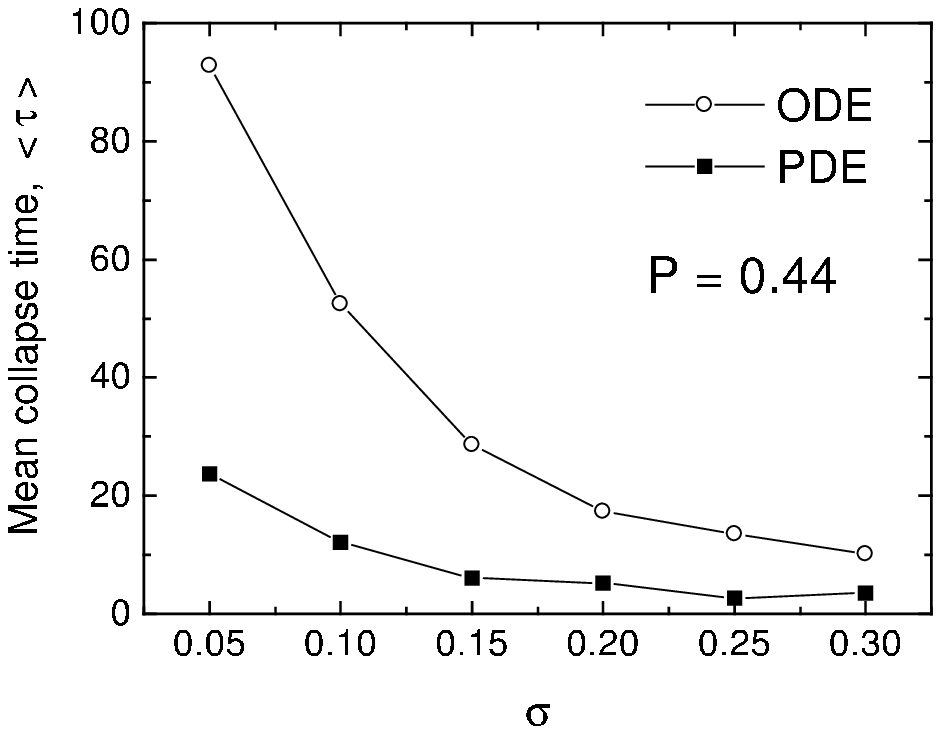}
\end{tabular}
\end{center}
\caption{
\label{fignum2}
Mean collapse time calculated from
stochastic PDE simulations (solid squares) and compared
with the corresponding stochastic ODE simulations (open circles).
Each mean is computed by averaging over a series of $100$ simulations.
Picture a: 
Mean collapse time as a function of $P$ for a white noise strength
$\sigma=0.3$.
Picture b:
Mean collapse time as a function of $\sigma$ for a nonlinear parameter
$P=0.44$ close to the critical value $P_c=0.459$.
}
\end{figure}

In a second step we have performed numerical simulations
of the GP equation (\ref{eq:schro1}) driven by a random Gaussian white
noise $\eta$ with zero-mean and autocorrelation
function $\EE[ \eta(t) \eta(t') ] = \sigma^2 \delta(t-t')$.
We do so by choosing randomly and independently the value of $\eta$ 
at each time step.
The mean
collapse time is calculated as an average over 100 realizations
of random paths along which the width of the condensate evolves
from the value corresponding to the minimum of the effective
potential $a_0$ until the value corresponding to its local
maximum $a_1$ (see Fig.~\ref{fig1}). The initial wave-form is selected as a
Gaussian with parameters predicted by the variational
approximation corresponding to the stationary state of the
condensate. 
Fig.~\ref{fignum2}a represents the collapse time for 
different values of the parameter $P$
which are not too close to the critical value $P_c$.
Comparison with the results from numerical 
simulations of the ODE (\ref{eq:ode1})
shows a very good agreement.
This demonstrates that the variational approach provides accurate
predictions for the behavior of the BEC.
for small non-linearity, and that the asymptotic analysis carried out
in Section \ref{sec:smallnl} holds true for the randomly driven GP
equation.

Finally, we have performed numerical simulations of the GP equation
(\ref{eq:schro1}) driven by a white noise $\eta$
with a nonlinear parameter $P$ very close to the critical
value $P_c=0.459$.
For near-critical values of the parameter $P$ the
Gaussian waveform was found to be not enough accurate. In this
case we employed the exact solution of the GP equation to initiate random
simulations. The exact solution (ground state) of the GP equation is
found by imaginary time-evolution method as described in
\cite{chiofalo}.
It is plotted in Fig.~\ref{fignum1}b. 
The results are plotted in Fig.~\ref{fignum2}b.
We can see that collapse in the perturbed PDE
occurs much earlier than in the ODE model.
This shows that the BEC in full GP equation is unstable against collapse at
near critical nonlinear parameter.
A small perturbation can drive the BEC to collapse through
fluctuations that are not captured by the variational approach.
Accordingly, we can state that the variational approach provides poor
predictions for the behavior of the BEC
for critical non-linearity.
Several reasons can explain the departure:
1) the Gaussian ansatz is not correct (see Fig.~\ref{fignum1}b).
2) the study of the ODE model shows that the important
parameter in the near-critical case is not
the value of $P$, but the value of the difference between
$P$ and $P_c$.
But the ODE does not capture the correct value of $P_c$,
so the error committed in the evaluation of the 
difference $P-P_c$ becomes very large when $P$ becomes close to $P_c$.
3) radiation effects become very important, in the sense that the 
waveform is strongly affected, even when the simulations 
are performed starting from 
the exact numerical waveform plotted in Fig.~\ref{fignum1}b,
so that we feel that it is useless to try to find a more suitable ansatz.
In this respect, one should add that this result is not surprising 
because it is well known in nonlinear optics
that the time-dependent variational approach fails to describe
the regime near the collapse \cite{kuznetsov,berge}.
Finally, it is necessary to mention that
the behavior of the gas close to collapse can be affected
by mechanisms that are not included in the GP equation,
such as inelastic two and three-body collisions 
\cite{edwards96,fedichev96}.

\section{Conclusion}
We have considered in this paper a condensate  trapped
by an external potential generated by a system 
of laser beams
in the case of a negative scattering length.
We have studied the stability of the metastable BEC
against small
fluctuations of the laser intensity.
We have shown that collapse of the BEC occurs whatever
the amplitude of the fluctuations after a time which is inversely
proportional to the integrated covariance of the 
autocorrelation function of the fluctuations of the laser intensity.
The statistical distribution of the collapse time has been computed.
The dependence of the mean collapse time with respect to the number atoms $N$
has been thoroughly analyzed.
We have shown that, for $N$ below the critical number of atoms
$N_c$, the mean collapse time decreases logarithmically with 
increasing $N$.
As a byproduct of the analysis we have shown that the variational approach
is very efficient for the analysis of the BEC for a number of atoms $N$
which is not too close to $N_c$, but we have seen that it completely fails
for $N$ close to $N_c$.

\section{Acknowledgments}
F. Kh. A. and B. B. B. are grateful to the Fund of fundamental
researches of the Uzbekistan
Academy of Sciences for the partial financial support.
B.B.B. also thanks the Physics Department of the
University of Salerno, Italy, for a research grant.

\end{document}